\documentclass{IEEEtran}
\IEEEoverridecommandlockouts

\usepackage[english]{babel}
\usepackage[T1]{fontenc}

\usepackage{amsmath}
\usepackage{graphicx}

\newcommand{\code}[1]{{\small\textsf{#1}}}

\newcommand{\hide}[1]{}
\newcommand{\anon}[2]{#2}

\title{How Developers Choose Names
\thanks{\anon{[Funding anonymized]}{
Dror Feitelson holds the Berthold Badler chair in Computer Science.
This research was supported by the ISRAEL SCIENCE FOUNDATION (grants no.\ 407/13 and 832/18).}
}}
\author{\anon{\IEEEauthorblockN{[Authors anonymized]}}{
\IEEEauthorblockN{Dror G. Feitelson ~~~~~ Ayelet Mizrahi ~~~~~ Nofar Noy\\
Aviad Ben Shabat ~~~~~ Or Eliyahu ~~~~~ Roy Sheffer}\\
\IEEEauthorblockA{Department of Computer Science\\
The Hebrew University of Jerusalem, 91904 Jerusalem, Israel}
}}

\begin{document}
\maketitle

\begin{abstract}
The names of variables and functions serve as implicit documentation and are instrumental for program comprehension.
But choosing good meaningful names is hard.
We perform a sequence of experiments in which a total of 334 subjects are required to choose names in given programming scenarios.
The first experiment shows that the probability that two developers would select the same name is low:
in the 47 instances in our experiments the median probability was only 6.9\%.
At the same time, given that a specific name is chosen, it is usually understood by the majority of developers.
Analysis of the names given in the experiment suggests a model where naming is a (not necessarily cognizant or serial) three-step process: (1) selecting the concepts to include in the name, (2) choosing the words to represent each concept, and (3) constructing a name using these words.
A followup experiment, using the same experimental setup, then checked whether using this model explicitly can improve the quality of names.
The results were that names selected by subjects using the model were judged by two independent judges to be superior to names chosen in the original experiment by a ratio of two-to-one.
Using the model appears to encourage the use of more concepts and longer names.
\end{abstract}

\begin{flushright}
\emph{And out of the ground the Lord God formed every beast\\ of the field, and every fowl of the air; and brought\\ them unto Adam to see what he would call them:\\
-- Genesis 2:19}\\
\end{flushright}

\section{Introduction}

The names of variables and functions are a major part of programs' source code.
In large open source projects about a third of the tokens are identifiers, and they account for about two thirds of the characters in the source code \cite{deissenboeck05}.
But the importance of names is not based only on their volume.
Their importance stems from the fact that they serve as implicit documentation, conveying to the reader the meaning of the code and the intent of the developer who wrote the code \cite{brooksr83,gellenbeck91}.
In fact, sometimes names are the \emph{only} documentation.
This is even advocated as part of the ``clean code'' approach, which states ``if a name requires a comment, then the name does not reveal its intent'' \cite{martin:clean}.

And indeed, it is generally agreed that meaningful names are instrumental aids for program comprehension \cite{gellenbeck91,blinman05,salviulo14}.
Programming courses therefore routinely require their students to ``use meaningful names''.
Books on programming may devote entire chapters to the issue of naming (e.g.\ McConnell's \emph{Code Complete} \cite[chap.\ 11]{mcconnell:cc} or Martin's \emph{Clean Code} \cite[chap.\ 2]{martin:clean}).
But there has been relatively little actual research on what ``meaningful names'' means \cite{liblit06}, and how to ensure that names are meaningful.

One of the reasons that naming is problematic is that names derive from natural language, typically from English.
But natural language is inherently ambiguous, including, for example, synonyms (different words that mean the same thing), polysemy (words with multiple meanings), and homonyms (different words with identical spelling).
As a result names chosen by one developer may not convey the expected meaning to another developer \cite{deissenboeck05}.
It is therefore interesting to see how names are chosen in practice, and whether the process of choosing names can be improved.

To investigate these issues we conducted a sequence of experiments, using a total of 334 students and professional developers as experimental subjects.
The core experiments present the subjects with several programming scenarios, and elicit names for key variables, data structures, and functions that are expected to be used in writing code for these scenarios.
The first and larger experiment checked spontaneous naming based on each subject's independent inclinations.
Based on the results of this experiment we developed a 3-step model of how names are constructed: (1) selecting the concepts to include in the name, (2) choosing the words to represent each concept, and (3) creating a name from these words.
The second experiment then checked what names are generated by subjects that have been introduced to this model explicitly, using exactly the same scenarios as the first experiment.
This was followed by an assessment by two independent judges of whether these names were superior over the names generated in the first experiment.

Our main contributions are as follows:
\begin{itemize}
    \item On the methodological level, we
    \begin{enumerate}
        \item Develop an experimental approach for studying how names are chosen in various programming scenarios without requiring the actual writing of voluminous code;
        \item Use bilingual experimental subjects to separate the description of scenarios from the coding, thereby reducing the danger that subjects would be guided toward specific names by wording used in the descriptions;
        \item Introduce head-to-head competitions where judges select the superior name from two candidates as a means to assess the relative quality of names in a given context.
    \end{enumerate}
    \item In terms of empirical results, we
    \begin{enumerate}
        \item Show that developers are indeed strongly influenced by the description of a scenario when they choose names in the context of that scenario;
        \item Show that the probability that two developers would choose the same name in the same situation is typically very low;
        \item Find that experienced developers tend to use longer names composed of more words;
        \item Characterize the structure of names and suggest a 3-step model that could explain how they are constructed.
    \end{enumerate}
    \item Finally, regarding practical implications, we
    \begin{enumerate}
        \item Demonstrate that developers can use the model of name construction to guide the process of choosing names;
        \item Find that names chosen using the model tend to contain more concepts, and are judged to be superior by a ratio of two-to-one.
    \end{enumerate}
\end{itemize}{}

\section{Related Work}

Practically all programming guidelines and textbooks state that variables should be given ``meaningful names''.
But spoken language is ambiguous, and people may use different words to refer to the same concepts and operations.
Furnas et al.\ studied spontaneous word choice for the commands of an interactive system, and found the variability to be surprisingly high: in every case two people favored the same term with a probability of less than 0.2 \cite{furnas87}.
In a related vein, Hindle et al.\ show that the vocabularies of different projects tend to be more diverse than the commonly used vocabulary in natural language \cite{hindle16}.
We perform an experiment similar to that of Furnas for variable and function names in code, and show that high variability exists in this context too.
To the best of our knowledge such a study has not been performed before.
We also introduce a novel methodology, where the description of the context is given in a different language (Hebrew) so as to reduce the effect on subjects.

Several studies have shown the importance of good names.
Rilling and Klemola show that code fragments with high identifier density may act as ``comprehension bottlenecks'' \cite{rilling03}.
Osman et al.\ demonstrated that names are crucial for understanding UML diagrams---without them there is no clue of what the different classes actually do \cite{osman12}.
Haiduc et al.\ found that when summarizing code, developers tend to include in the summary practically all the terms that appear in method names, and the vast majority of terms in parameter types \cite{haiduc10}.
In an ethnographic study, Salviulo and Scanniello found that experienced developers prefer to rely on variable names rather than on comments for comprehension and maintenance \cite{salviulo14}.

Conversely, some studies have shown the detrimental effects of bad names.
Butler et al.\ found that flawed identifier names are associated with low quality code \cite{butler10}.
Arnaoudova et al.\ introduced the notion of ``linguistic antipatterns'' where names are misleading \cite{arnaoudova16}.
Problematic situations include mismatch of type, number, or behavior (e.g.\ a `set' method that returns a value).
Avidan and Feitelson have shown that misleading identifier names are worse than meaningless names like consecutive letters of the alphabet, as they may lead to errors in program comprehension \cite{avidan17}.
These results emphasize the importance of identifier naming, and indicate that guidance on naming identifiers may be beneficial.

Perhaps the most studied aspect of variable names is their length, and in particular the possible use of abbreviations.
Binkley et al.\ suggest that long names tax memory, so a limited length and a limited vocabulary are preferable \cite{binkley09c}.
Interestingly, even single-letter names may convey meaning, provided the right letter is chosen \cite{beniamini17,swidan17}.
Several studies have found that abbreviations do not have a significant effect.
Lawrie et al.\ found that in many cases abbreviations are just as understandable as longer names \cite{lawrie06}.
Scanniello et al.\ showed that full names did not help novice programmers find and fix faults \cite{scanniello13}.
However, other studies claimed that abbreviations are detrimental to comprehension.
Takang et al.\ find that full names are better than abbreviations \cite{takang96}, and Hofmeister et al.\ and Schankin et al.\ claim that comprehension is faster with full and descriptive names \cite{hofmeister17,schankin18}.

Turning to suggestions for improving variable names,
Caprile and Tonella suggest standardization of variable names using a lexicon of concepts and syntactic rules for arranging them \cite{caprile00}.
Dei{\ss}enb\"ock and Pizka stress the need for concise and consistent naming \cite{deissenboeck05}.
Binkley et al.\ suggest rules for enhancing the information expressed by  field names \cite{binkley11}.
Such approaches can mesh nicely with our model of name construction.

The possibility of tool support for selecting names has also been discussed.
Allamanis et al.\ noted that automatic suggestion of function or class names is harder than suggesting variable names, because of the need to describe abstract functionality \cite{allamanis15}.
Conversely, Liu et al.\ have shown that in many cases the contents of the body of a function can be used to construct a good function name \cite{liu19}.
Raychev used machine learning on ``big code'' to predict variable names, achieving 62\% accuracy \cite{raychev15}, and Alon et al.\ exploit the code structure to predict method names \cite{alon19}.
Our focus, in contradistinction, is on helping developers think about naming.

At a deeper level of analysis, it is very interesting to consider the interaction between code comprehension and cognitive processes in the brain \cite{liblit06,storey06}.
Brooks has theorized that names support top-down comprehension based on forming hypotheses regarding what the code does \cite{brooksr83}.
Groundbreaking work by Siegmund et al.\ using fMRI brain imaging has shown that understanding source code activates parts of the brain related to language processing \cite{siegmund14}.
In later work, Siegmund et al.\ found evidence for using semantic chunking when performing bottom-up comprehension tasks \cite{siegmund17}.
Fakhoury et al.\ combined the fNIRS brain-imaging technique with eye tracking to show that linguistic anti-patterns in the code correlate with cognitive load as measured directly in the brain \cite{fakhoury19b}.
Works like these contribute significantly to our understanding and appreciation of the importance of naming.

\section{Experiment on Name Selection}

Our general goal is to better understand the issue of meaningful names and their role in program comprehension.
This starts with basic science: understanding how names are formed as part of understanding why naming is hard and sometimes problematic.
Our first experiment concerns how developers choose names.
It employs a survey, given to professional developers and computer science students, to see what names they choose in various scenarios, and how they understand given names.

\subsection{Research Questions}

We concretize the above focus using the following research questions:
\newcounter{rqnum}
{\renewcommand{\theenumi}{RQ\arabic{enumi}}
\begin{enumerate}
\item\label{rq:same}
What is the probability that different developers will select the same name in the same scenario?

To answer this question, we also need to consider two sub-questions:
\renewcommand{\theenumii}{.\arabic{enumii}}
\renewcommand{\labelenumii}{\theenumi\theenumii)}
\begin{enumerate}
\item\label{rq:prime} Is name choice affected by the wording used in the description of the scenario?
\item\label{rq:exp} Is name choice affected by demographics (e.g.\ experience or sex)?
\end{enumerate}

\item\label{rq:understand} 
How well do developers understand names chosen by others?

\item\label{rq:struct} What is the structure of chosen names?
In particular, what is the distribution of name lengths, and to what degree are multi-word phrases used?
\setcounter{rqnum}{\value{enumi}}
\end{enumerate}}

\subsection{Methodology}

We use a survey on the Internet to present scenarios and ask how relevant constructs would be named or how named constructs are understood.
This is an unusual experiment because we are primarily interested in the variability between subjects.

\subsubsection{Survey Structure}

The survey was structured as a sequence of independent scenarios.
Each scenario lays out a situation or a problem, followed by several focused questions.
Question types include:
\begin{itemize}
\item Suggest a name for a variable, constant, or data structure (16 questions).
\item Suggest a signature for a function, including its name and its parameters (5 questions).
\item Interpret a given variable  or function parameter name (9 questions).
\item Interpret a given function signature (6 questions).
In some cases this included writing a line of pseudo-code to remove ambiguity.
\end{itemize}

A major problem in studying spontaneous naming is that the description of the context and the question itself necessarily use words.
Being exposed to these words makes then more accessible, and therefore subjects will tend to use words from the description in the names they create.
Thus providing the description undermines the spontaneity we are trying to characterize.

We reduce the accessibility problem by using multilingual subjects.
In particular, our subjects are typically native Hebrew speakers who are also fluent in English.
We can then provide the description in Hebrew, and this is expected to have a much smaller effect on the names chosen in English (which is universally used in programming).
As a control other subjects are given the description in English, to establish whether the accessibility effect indeed exists.

In all, 11 scenarios were defined, most of them including 3--6 questions.
8 of the scenarios had both Hebrew and English versions (see brief synopsis in Table \ref{tab:2hit}).
The 3 scenarios that had only a Hebrew version did not require respondents to produce names, but asked only about their understanding of given names.
2 of the multi-version scenarios had 2 separate English versions in addition to the Hebrew version, using different wording in English.
In total 121 questions were needed for all the versions of all the questions.
All experimental materials, including a full listing of the survey questions, are available for additional study and reproduction (see link at end).

\subsubsection{Survey Execution}

The survey was conducted using the Qualtrics web-based platform.
Subjects were presented with 6 randomly selected scenarios in a random order.
For each scenario, one version was selected at random; thus no subject saw multiple versions of the same scenario.
No personally identifying data was collected.
None of the questions were compulsory; respondents could skip questions or stop without completing the survey.
In such cases, we use only the part that was completed.

The survey was conducted in May 2018 in two phases:
first with professional developers and then with students.
Developers were recruited via personal contacts with colleagues in the high-tech industry
and requesting them to propagate it.
Students were recruited in student labs by handing out the link to the survey together with a chocolate treat.
Altogether, 234 respondents answered at least one question.

The survey started with a few demographic questions.
75\% of the respondents were male, and 25\% female.
The most common ages were in the range 24--27, but the mean age was 27.9 years.
Experience ranged from 0 to 29 years, with an average of 5.8 years.
70\% identified themselves as students in computer science or engineering.

\subsubsection{Analysis of Results}

Importantly, the survey questions were open text and not multiple-choice, so as to give respondents the opportunity to select names and express themselves.
This makes the analysis of the results much harder, as we need to identify cases where different responses actually mean the same thing.
We automated part of this process, but much of the analysis required going over the responses manually.

In order to compare names, we use a normalization procedure.
First, we identify suspicious cases and divert them for manual inspection.
We define suspicious names as
\begin{itemize}
\item having more that 30 characters.
\item containing non alphanumeric or underscore characters.
\end{itemize}
In many cases, suspicious names were indeed not names but comments by the experimental subjects (such as ``I don't know what to do here'').
Such answers were removed from further consideration.

\begin{figure*}\centering
\includegraphics[width=\textwidth]{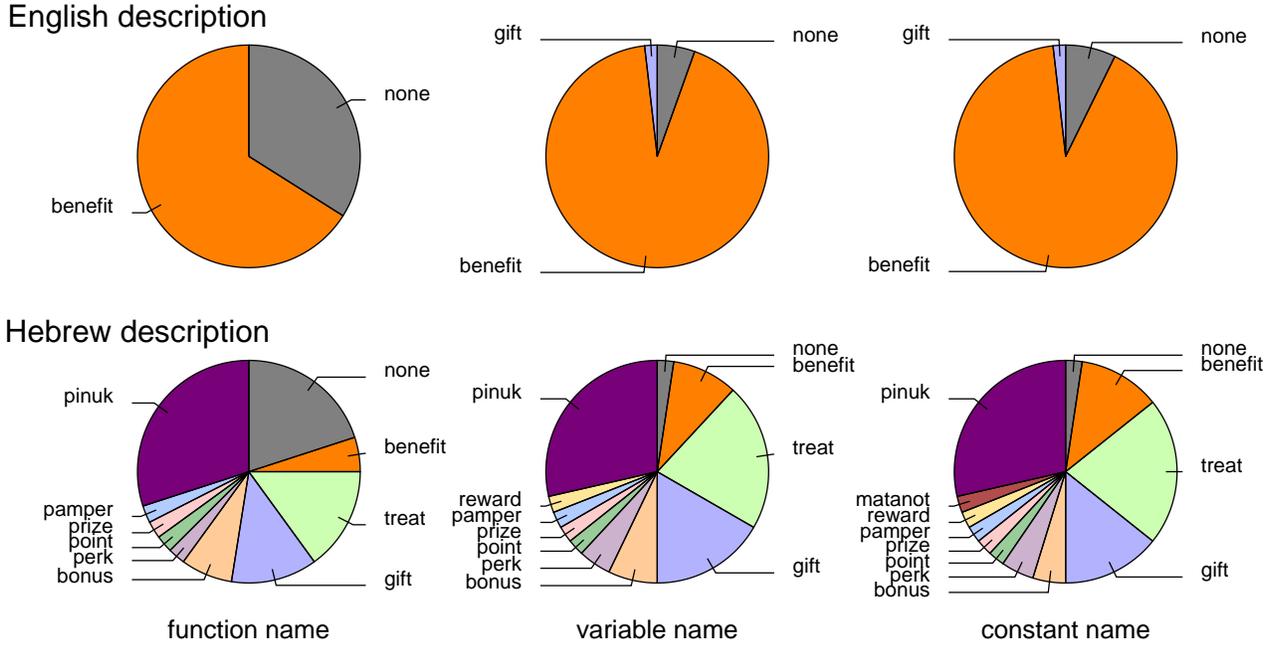}
\vspace*{-8mm}
\caption{\label{fig:prime-benefits}\sl
Choice of words for expressing the concept of ``benefit'' when the description was in English and explicitly used ``benefit'', as opposed to a Hebrew description that used the word ``pinuk''.
Variations on the same word (spelling or plurality) are shown together.
``None'' means that the concept was not represented in the name.}
\end{figure*}

Second, names were normalized to lower case with words separated by \_.
The normalization first partitions the name into a sequence of words, using a regular expression that recognizes camelCase and a variety of separators (-, \_, ., and white space).
Each word is then converted to lowercase.
Finally the words are rejoined using \_.
Then names with a Levenshtein distance $\leq 2$ are identified.
This means that one name can be transformed into the other by up to 2 single-letter edits (insertion, deletion, or substitution) \cite{levenshtein66}.
This allows us to avoid variability that actually reflects only spelling errors, e.g.\ if someone wrote ``bord'' instead of ``board''.
All such corrections were verified manually.

Questions about function signatures were handled similarly.
We use a regular expression for identifying a general C-like function signature (optional return type, name, and parentheses with a comma-separated list of arguments).
Each name is then displayed interactively to the analyst to identify its meaning relative to the question.
This step is necessary for function parameters because experimental subjects may present the parameters in arbitrary order.
The answers are remembered in case the same parameter name is used again later.

The hardest questions to analyze are those that require the experimental subjects to say
what they think some name means.
These answers were clustered based on our best understanding into clusters of answers that essentially say the same.
In order to ensure that our classification is not too subjective, we recruited an external analyst to check a sample of the classifications.
In nearly all cases, there was agreement.

\begin{figure*}\centering
\includegraphics[width=\textwidth]{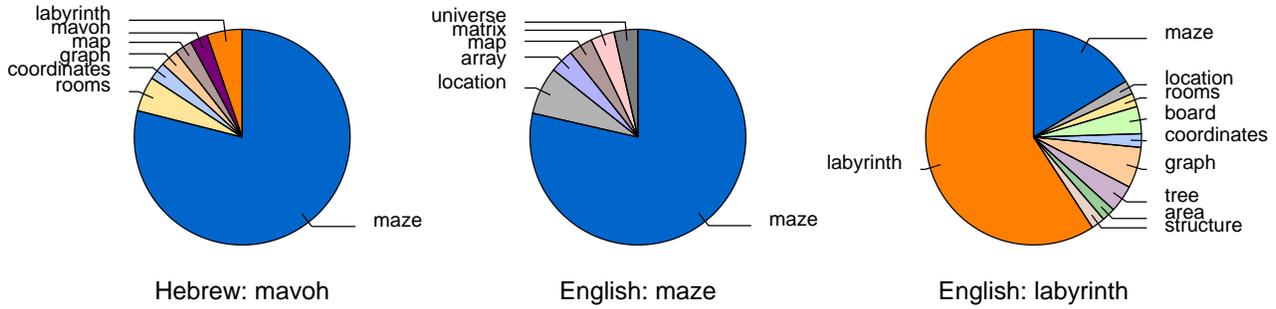}
\vspace*{-8mm}
\caption{\label{fig:prime-maze}\sl
Choice of words used in naming a data structure describing a maze when the English description used ``maze'' or ``labyrinth'', and when the description was given in Hebrew using ``mavoh''.
Variations on the same word (spelling or plurality) are shown together.
Words not related to the maze (e.g.\ ``descriptor'') are excluded.}
\end{figure*}

\subsection{Results}

As noted, our survey included 121 questions answered by 234 subjects (each answering only a subset), in many cases leading to a wide distribution of results.
We cannot show all of these results here.
Rather, we focus on the main trends and on interesting examples.
The full raw results are available as part of the experimental materials.

\subsubsection{The Effect of Accessibility}
\label{sect:prime}

Many of the questions involved choosing names for variables, data structures, or functions.
As noted above, a potential problem with studying this is the possibility of an accessibility effect: by describing the scenario in which a name needs to be chosen, we expose the subjects to specific words that are candidates for being used in the name.
To check this effect and answer \ref{rq:prime} we used bilingual subjects, and provided some of them with a description in Hebrew rather than in English.

An example of the results obtained is shown in Fig.\ \ref{fig:prime-benefits}.
The scenario was the calculation of benefits accrued by using a credit card, where a point is awarded for each 2000 Shekels charged up to a maximum of 4 points per month.
Three of the questions were to name a function used to determine whether a customer has additional benefit points available this month, to name a variable storing the number of benefits available, and to name the constant 4.
The names suggested in each case were varied, but most of them contained
the concept of ``benefits''.
When presented with an English description, practically all the subjects actually used the word ``benefit''.
The Hebrew description used the word ``pinuk'', and this transliteration or the translation to ``treat''  were the most common words used.
But subjects also used ``gift'', ``benefit'', ``bonus'', and some other terms.

Another example is shown in Fig.\ \ref{fig:prime-maze}.
In this case the scenario was a mouse looking for cheese in a maze, and the question was to name the data structure representing the maze.
This time two separate English descriptions were used: one using the word ``maze'', and the other using ``labyrinth''.
The results show a strong accessibility effect, including when the descriptions was in Hebrew, where the Hebrew word ``mavoh'' was nearly universally translated to ``maze''.
The most diverse results were obtained for the English description using ``labyrinth'', including quite a few subjects who chose to simplify and use ``maze''.

\begin{figure}\centering
\includegraphics[width=0.7\columnwidth]{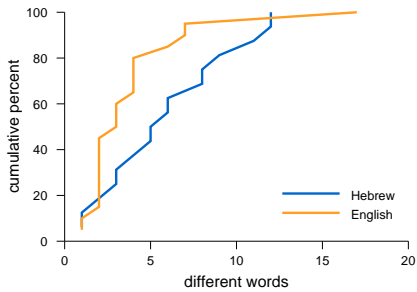}
\vspace*{-4mm}
\caption{\label{fig:words-dist}\sl
Distributions of the number of different words used for the main concept when naming variables.}
\end{figure}

In other cases the differences were not as extreme as in these two cases.
The distributions of the number of different words used for the main concept in the 36 questions on naming variables are shown in Fig.\ \ref{fig:words-dist}.
Comparing the equivalent Hebrew and English questions, in 9 cases responses to the Hebrew one used more different words, in 4 cases the English version led to using more words, in 2 cases the same number was used, and in 1 the results were mixed (there were 2 English versions, and the Hebrew result was in the middle).
Thus the Hebrew descriptions indeed typically led to a larger diversity in word choice, or conversely, the English descriptions led to more focus on fewer words.
We conclude that accessibility may be an issue, and using Hebrew descriptions (or another foreign language) may reduce this effect.

\subsubsection{Name Structure and Length}

\begin{figure}\centering
\includegraphics[width=0.9\columnwidth]{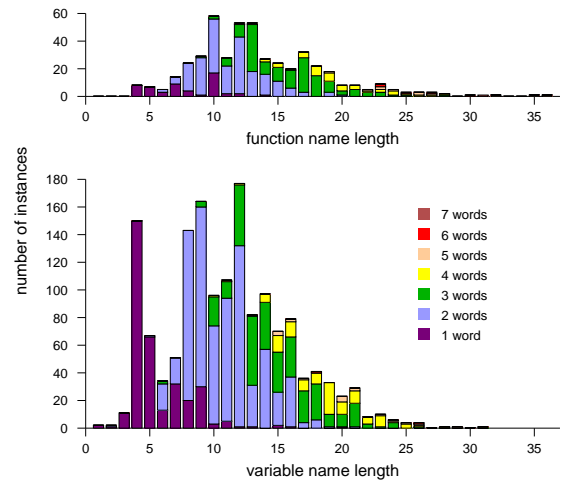}
\vspace*{-4mm}
\caption{\label{fig:name-len}\sl
Histogram of name lengths for variables and functions.
Underscores separating words are not counted.}
\end{figure}

Names are often formed be combining multiple words.
Moreover, an empirical study performed by Holzmann identified a trend where function and variable names are becoming longer \cite{holzmann16}.
To answer \ref{rq:struct}, we now characterize the given names in terms of their structure.

The distribution of the name lengths is shown in Fig.\ \ref{fig:name-len}.
Function names tend to be slightly longer than variable names (2.43 characters longer on average, p<0.0001 using t test).
The distribution of variable name lengths is bimodal.
Names that are composed of one word are typically 4--5 characters long, and rarely exceed 9 characters.
Names that are composed of 2 or more words are typically 8--16 characters long.
Longer names are typically composed of 3 or more words.

As indicated in the figure, the vast majority of names given were composed of multiple words (78\% of variable names and 89\% of function names).
This also reflects the number of different concepts included in names, which is discussed later.

\ref{rq:exp} concerned the interesting question of whether demographic variables interact with name lengths.
We first checked the effect of experience.
Following Falessi et al.\ \cite{falessi18} this was not a simple classification into students and professionals; rather, we define experienced developers to be those with at least 5 years work experience (even if they are currently also students), and inexperienced ones to be 1st or 2nd year BSc students provided they have at most 2 years experience.
In our dataset, there were 100 experienced developers according to this definition, 43 inexperienced students, and 75 participants that fell in between (an additional 12 did not provide all the required data).

\begin{figure}\centering
\includegraphics[width=0.7\columnwidth]{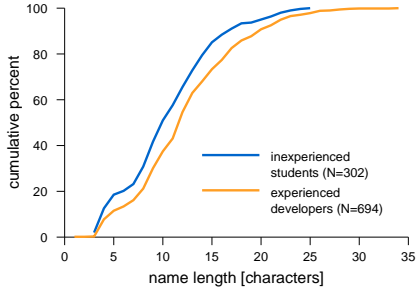}
\vspace*{-4mm}
\caption{\label{fig:exp-len-char}\sl
The cumulative distribution of name lengths given by inexperienced and experienced developers.
$N$ refers to the number of names given.}
\end{figure}

Fig.\ \ref{fig:exp-len-char} shows the distributions of name lengths chosen by inexperienced vs.\ experienced developers.
The result is that the distribution of names chosen by experienced developers dominates the distribution of names chosen by inexperienced students.
This means that for every name length, the probability for experienced developers to use a name of this length or more is higher than the corresponding probability for inexperienced ones.
The difference is that on average experienced developers use names that are 1.75 characters longer (p<0.0001 using t test).
This difference is largely the result of experienced developers using names with more words (and by implication, more concepts): 26.2\% of the names by inexperienced developers have only 1 word and the average length is 2.03 words, as opposed to only 16.7\% and 2.29 words for experienced ones.

We also checked the effect of sex, and found no difference in the lengths of names given by males and females: the distributions of name lengths given by the two sexes were practically on top of each other.

\begin{table*}\centering
\caption{\label{tab:2hit}\sl
Results of name reuse in all versions of questions concerned with giving names.
Note that these results are for the complete names, not for concepts.
\emph{$N$}: number of answers to this question version;
\emph{dif}: number of different names given;
\emph{div=dif/$N$}: diversity of names;
\emph{max}: maximal answers giving the same name;
\emph{Pmax=max/$N$}: probability of most popular name (focus);
\emph{P2hit}: probability of two respondents using the same name.
}
\begin{tabular}{@{}lllcccccc@{}}
\hline
\emph{Scenario} & \emph{Question} & \emph{Version} & \emph{$N$} & \emph{dif} & \emph{div} & \emph{max} & \emph{Pmax} & \emph{P2hit} \\
\hline
benefits	& function checking if balance of benefits is positive
		&  Eng 	& 54	& 38	& 0.703	& 8	& 0.148	& 0.0459 \\
card&	&  Heb 	& 41	& 40	& 0.975	& 2	& 0.048	& 0.0255 \\

	& constant with value 4 (max benefits per month)
		&  Eng 	& 55	& 26	& 0.472	& 14	& 0.254	& 0.1266 \\
	&	&  Heb	& 44	& 35	& 0.795	& 3	& 0.068	& 0.0340 \\

	& constant with value 2000 (shekels per benefits point)
		&  Eng 	& 55	& 45	& 0.818	& 5	& 0.090	& 0.0300 \\
	&	&  Heb 	& 44	& 42	& 0.954	& 2	& 0.045	& 0.0247 \\

	& variable with entitled benefits this month
		&  Eng 	& 55	& 44	& 0.800	& 3	& 0.054	& 0.0267 \\
	&	&  Heb 	& 44	& 41	& 0.931	& 2	& 0.045	& 0.0258 \\

\hline
elevator	& variable with requested floor
		&  Eng 	& 41	& 26	& 0.634	& 6	& 0.146	& 0.0577 \\
	&	&  Heb 	& 49	& 28	& 0.571	& 8	& 0.163	& 0.0645 \\

	& variable with current elevator location
		&  Eng 	& 41	& 17	& 0.414	& 21	& 0.512	& 0.2861 \\
	&	&  Heb 	& 49	& 19	& 0.387	& 21	& 0.428	& 0.2161 \\

	& variable with number of floors to move
		&  Eng 	& 41	& 29	& 0.707	& 5	& 0.121	& 0.0505 \\
	&	&  Heb 	& 49	& 37	& 0.755	& 5	& 0.102	& 0.0362 \\

	& variable with state of elevator doors
		&  Eng 	& 40	& 18	& 0.450	& 12	& 0.300	& 0.1312 \\
	&	&  Heb 	& 49	& 17	& 0.346	& 11	& 0.224	& 0.1303 \\

\hline
file	& function checking if there is enough disk space to extend a file
		&  Eng 	& 44	& 35	& 0.795	& 4	& 0.090	& 0.0361 \\
system&	&  Heb 	& 54	& 47	& 0.870	& 3	& 0.055	& 0.0240 \\

	& field in file object describing file size
		&  Eng 	& 44	& 11	& 0.250	& 21	& 0.477	& 0.3336 \\
	&	&  Heb 	& 56	& 7	    & 0.125	& 26	& 0.464	& 0.4164 \\

\hline
bubble	& constant specifying work hours per week
		&  Eng 	& 42	& 36	& 0.857	& 3	& 0.071	& 0.0328 \\
gum	&	&  Heb 	& 43	& 35	& 0.813	& 5	& 0.116	& 0.0416 \\

factory& variable holding hourly wage during overtime
		&  Eng 	& 42	& 24	& 0.571	& 10	& 0.238	& 0.0941 \\
	&	&  Heb 	& 43	& 37	& 0.860	& 3	    & 0.069	& 0.0319 \\

\hline
ice cream	& function calculating how many sandwiches can be produced
		    &  Eng 	& 41	& 36	& 0.878	& 2	& 0.048	& 0.0303 \\
sandwich&	&  Heb 	& 47	& 44	& 0.936	& 2	& 0.042	& 0.0239 \\

\hline
maze	& variable with location of the cheese today
		&  Eng-maze 	& 27	& 18	& 0.666	& 8	    & 0.296	& 0.1193 \\
	&	&  Eng-labyrinth& 44	& 20	& 0.454	& 13	& 0.295	& 0.1363 \\
	&	&  Heb       	& 35	& 24	& 0.685	& 5	    & 0.142	& 0.0628 \\

	& data structure tracking where the mouse has visited
		&  Eng-maze 	& 27	& 13	& 0.481	& 9	    & 0.333	& 0.1550 \\
	&	&  Eng-labyrinth& 43	& 21	& 0.488	& 13	& 0.302	& 0.1346 \\
	&	&  Heb          & 35	& 26	& 0.742	& 7	    & 0.200	& 0.0693 \\

	& data structure describing the maze
		&  Eng-maze 	& 27	& 10	& 0.370	& 17	& 0.629	& 0.4128 \\
	&	&  Eng-labyrinth& 43	& 16	& 0.372	& 21	& 0.488	& 0.2742 \\
	&	&  Heb 	        & 35	& 8	    & 0.228	& 28	& 0.800	& 0.6457 \\

\hline
mine-	& function calculating game's difficulty level
		    &  Eng 	& 45	& 25	& 0.555	& 8	& 0.177	& 0.0706 \\
sweeper &	&  Heb 	& 52	& 25	& 0.480	& 7	& 0.134	& 0.0680 \\

	& variable with game's time
		&  Eng 	& 47	& 22	& 0.468	& 11 & 0.234	& 0.1027 \\
	&	&  Heb 	& 51	& 29	& 0.568	& 7	 & 0.137	& 0.0588 \\

	& data structure indicating mine or number of adjacent mines
		&  Eng 	& 43	& 33	& 0.767	& 5	& 0.116	& 0.0416 \\
	&	&  Heb 	& 52	& 42	& 0.807	& 6	& 0.115	& 0.0340 \\

\hline
tic-tac-toe	& function to display the game board
		&  Eng-board 	& 40	& 15	& 0.375	& 16 & 0.400	& 0.1950 \\
	&	&  Eng-grid 	& 27	& 15	& 0.555	& 4	 & 0.148	& 0.0864 \\
	&	&  Heb 	        & 28	& 16	& 0.571	& 5	 & 0.178	& 0.0943 \\

	& data structure describing current state of game board
		&  Eng-board 	& 40	& 25	& 0.625	& 8	 & 0.200	& 0.0737 \\
	&	&  Eng-grid 	& 26	& 16	& 0.615	& 4	 & 0.153	& 0.0857 \\
	&	&  Heb 	        & 28	& 12	& 0.428	& 16 & 0.571	& 0.3443 \\

\hline
\end{tabular}
\end{table*}

\subsubsection{The Probability of Using the Same Name}

The results in Sect.\ \ref{sect:prime} above are actually more about the naming of concepts than the naming of variables.
Variable and function names tend to be more varied than the words used to denote a concept, for two reasons.
First, there can be variations on using the same word, such as using it in the singular or in plural.
Second, variable and function names are often actually multi-word phrases, and these components can be strung together in different ways.
As a result, complete names given by different subjects are often unique.

We now return to \ref{rq:same}.
Table \ref{tab:2hit} summarizes the results regarding the variability of names.
The distribution of names given as answers to a given question can be characterized by two related attributes:
\begin{itemize}
\item \emph{diversity}: the degree to which they are diverse.
The number of different names is shown in column \emph{dif}.
The diversity is defined by normalizing this by the total number of responses received (column \emph{div}).
\item \emph{focus}: The degree to which names are focused.
Column \emph{max} gives the number of times that the most popular name was used.
The fraction of responses that used the same most popular name is shown in column \emph{Pmax}.
\end{itemize}

\begin{figure}\centering
\includegraphics[width=0.7\columnwidth]{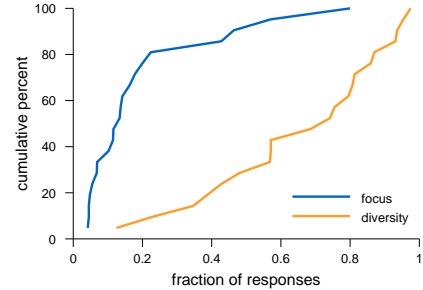}
\vspace*{-4mm}
\caption{\label{fig:focus-divers}\sl
Cumulative distribution of focus and diversity of names given in the Hebrew questions.}
\end{figure}

Fig.\ \ref{fig:focus-divers} shows the distribution of the diversity and focus.
We use the results of only the Hebrew versions of questions, to reduce the accessibility effect.
As can be seen, in 80\% of the questions the most popular name was used in only 5-20\% of the answers.
And in half of the questions the number of different names given was equivalent to at least 3/4 of the number of answers.
As may be expected, focus and diversity are inversely correlated (Pearson correlation of $-$0.817 on 21 Hebrew questions).

The final column of Table \ref{tab:2hit} estimates the probability of two developers to use the same name (answering \ref{rq:same}).
Note that this is not the probability that any other respondent happened to give the same name, as this obviously depends on the number of respondents (the birthday paradox).
Rather, we are interested in the probability that a specific developer reading the code would have chosen the same name as the specific developer who wrote it in the first place.
Assuming we observe $k$ distinct names, and using the relative popularity of name $i$ as an estimate for the probability of choosing it $p_i$, the probability of two developers choosing the same name is simply%
\footnote{The same formula is used in similar circumstances in other fields, to wit the Simpson diversity index in ecology, where $p_i$ represents the proportional abundance of species $i$, and the Harfindahl Index measuring market concentration, where $p_i$ is the market share of firm $i$.}
\begin{equation}\label{eq:p2hit}
    P2hit = \sum_{i=1}^k p_i^2
\end{equation}
In our results this ranges from 2.5\% to 64.6\%, with a median of 6.9\%.

\subsubsection{Understanding Names}

\begin{table*}\centering
\caption{\label{tab:understanding}
Results of how variable names or function signatures are understood.
\emph{$N$}: number of answers; fractional numbers indicate that an answer included two options.
Percentages may not sum to 100 when irrelevant answers (e.g.\ ``unclear'') were ommitted.}
\begin{tabular}{lllrrrr}
\hline
    &   &   & \multicolumn{2}{c}{Hebrew} & \multicolumn{2}{c}{English} \\
\emph{Scenario} & \emph{Code} & \emph{Interpretation} & \emph{N} & \emph{percent} & \emph{N} & \emph{percent} \\
\hline
file    & \code{arrangeFilesByName(files)}
            & returns ordered list of files & 53    & 96.4\%    & 44    & 97.8\% \\
system  &   & show sorted list of files     & 1     & 1.8\%     & 0     & --     \\
        &   & sort in place                 & 0     & --        & 1     & 2.2\%  \\[1mm]
        & role of \code{files} parameter
            & files to sort                 & 23    & 88.5\%    & 24    & 92.3\% \\
        &   & path/reference to files       & 2     & 7.7\%     & 2     & 7.7\%  \\
        &   & file sizes                    & 1     & 3.8\%     & 0     & --     \\
\hline
bubble  & \code{pay(hours, rate)}
            & returns amount to pay         & 19.5  & 45.3\%    & 30    & 68.2\% \\
gum     &   & transfers funds to employee   & 22.5  & 52.3\%    & 14    & 31.8\% \\
factory &   & set employee's rate \& hours  & 1     & 2.3\%     & 0     & --     \\[1mm]
        & role of \code{hours} parameter
            & number of work hours          & 38    & 90.5\%    & 37    & 86.0\% \\
        &   & work hours since last paid    & 0     & --        & 1     & 2.3\%  \\
        &   & work hours per month          & 3     & 7.1\%     & 2     & 4.7\%  \\
        &   & work hours per week           & 1     & 2.4\%     & 0     & --     \\
        &   & work hours per pay period     & 1     & 2.4\%     & 2     & 4.7\%  \\
\hline
ice cream & \code{profit(units, cost, price)}
            & calculate profit from selling & 46    & 95.8\%    & 40    & 97.6\% \\
sandwich&   & calculate profit from buying  & 1     & 2.1\%     & 0     & --     \\
        &   & total cost given unit cost    & 1     & 2.1\%     & 0     & --     \\
        &   & profit per unit               & 0     & --        & 1     & 2.4\%  \\[1mm]
        & role of \code{units} parameter
            & sandwiches produced/sold      & 44    & 93.6\%    & 33    & 91.7\% \\
        &   & amount of each ingredient     & 3     & 6.4\%     & 3     & 8.3\%  \\[1mm]
        & role of \code{cost} parameter
            & producing cost per unit       & 35.5  & 77.2\%    & 34.5  & 93.2\% \\
        &   & total producing cost          & 5.5   & 12.0\%    & 2.5   & 6.8\%  \\
        &   & producing cost                & 5     & 10.9\%    & 0     & --     \\[1mm]
        & role of \code{price} parameter
            & selling price per unit        & 43    & 97.7\%    & 38    & 100\% \\
        &   & total seling price            & 1     & 2.3\%     & 0     & --    \\
\hline
mine-   & \code{expose(row, col)}
            & expose a tile                 & 29    & 58.0\%    & 24    & 55.8\% \\
sweeper &   & expose tile \& execute game rules& 10 & 20.0\%    & 13    & 30.2\% \\
        &   & expose tile \& return its value & 2   & 4.0\%     & 0     & --     \\
        &   & update the GUI                & 1     & 2.0\%     & 1     & 2.3\%  \\
        &   & execute game logic for move   & 1     & 2.0\%     & 0     & --     \\
        &   & return tile value (has mine)  & 5     & 10.0\%    & 2     & 4.7\%  \\
        &   & return list of neighboring mines & 0  & --        & 1     & 2.3\%  \\
\hline
tic-tac-toe & \code{make\_turn(int row, int col)}
            & update board with mark at position& 27& 93.1\%    & 63    & 92.6\% \\
        &   & play the turn                 & 2     & 6.9\%     & 2     & 2.9\%  \\
        &   & move to given position        & 0     & --        & 1     & 1.5\%  \\
        &   & initialize for input          & 0     & --        & 1     & 1.5\%  \\[1mm]
        & role of \code{row} parameter
            & row index to be updated       & 29    & 100\%     & 66    & 100\%  \\[1mm]
        & role of \code{col} parameter
            & column index to be updated    & 29    & 100\%     & 66    & 100\%  \\
\hline
add     & return value of \code{add([1,2,3], [4,5,6])}
            & [5,7,9]                       & 78    & 70.9\%    & & \\
        &   & [1,2,3,4,5,6]                 & 23    & 20.9\%    & & \\
        &   & [6,15]                        & 1     & 0.9\%     & & \\
        &   & 21                            & 1     & 0.9\%     & & \\
        &   & error / Boolean               & 5     & 4.5\%     & & \\
\hline
resize  & apply \code{resize(factor)} to an image
            & resize the image              & 64    & 61.0\%    & & \\
        &   & enlarge the image             & 32    & 30.5\%    & & \\
        &   & shrink the image              & 3     & 2.9\%     & & \\
        &   & change pixel values           & 1     & 1.0\%     & & \\
\hline
\end{tabular}
\end{table*}

As opposed to active naming, which often led to very diverse results as reported above, interpreting given names led to more uniform results. This answers \ref{rq:understand}.
In these questions respondents were asked to interpret given names, function signatures, or function parameters.
In most cases nearly all of them agreed on the meaning.
However, they sometimes differed regarding technical details.
The results are shown in Table \ref{tab:understanding}.

As an example, we asked about the function \code{pay(hours, rate)} in a factory setting.
The consensus was that this refers to paying for work done.
But 49 thought that it calculates the sum to pay, while 36 others thought it actually transfers the funds.
Only one mentioned both possibilities, and one came up with a third possible meaning, suggesting that the function defines an employee's rate and required hours.

Another example concerned the function \code{arrangeFilesByName(files)}.
When asked what the function does, 98 of 100 agreed that it sorts a list of files, with 72 saying it is ordered lexicographically, and 10 specifying ascending order.
When asked about the role of the parameter, all but one agreed that it represents the files to sort.
But when asked about the expected return value,
opinions differed.
The most common response (69 respondents) was that the function returns a sorted list of files, but there were differences of opinion on whether this was a list of file names, file references, or file indices.
29 said it could be \code{void}, with 15 specifically mentioning in-place sorting.
7 expected a Boolean or numeric return code
(one suggested the number of files reordered).

These results are encouraging in the sense that even if different developers tend to come up with different names for things, they generally understand names chosen by others.
However, this may hide different assumptions regarding the technical details underlying this general understanding,
which can lead to bugs.

Anticipating that understanding may be affected by ambiguity, we specifically designed two of the questions about the meaning of names to be ambiguous, but without being unrealistic.
And indeed survey respondents understood them in different ways.

The first ambiguous question concerned a library function \code{add(a,b)}, and asked what would be the result of calling \code{add([1,2,3],[4,5,6])}.
The most common response by far was to understand the add operation applied to vectors as an element-wise addition, leading to a result of \code{[5,7,9]}
(some making an arithmetic error).
Others understood the add operation as a concatenation, and said the result would be \code{[1,2,3,4,5,6]}.
Only four noted the ambiguity and gave both options.

The second ambiguous question concerned a function with the signature \code{resize(factor)} applied to an image.
When asked to describe what this function does, 58\% of respondents just said it resizes the image.
But 30\% specifically said it \emph{enlarges} the image, and three said it \emph{shrinks} it.
In addition, we requested respondents to write the line of code that updates the width of the image.
The result was that 60\% multiplied the width by the factor, while 5\% divided it.
An additional 4\% multiplied by the square root of the factor, indicating that they understood the factor as applying to the area instead of to the individual dimensions.
An unanticipated popular response was to simply set the width to the factor (22\% of respondents).

\section{Analysis and Model of Name Formation}
\label{sect:model}

A model is the concise conceptual description of a system or process.
Scientific models are used to improve understanding of observed phenomena by focusing attention on salient features and explaining the relationships between them.
We therefore analyzed the names produced by experimental subjects in the experiment described above, in order to try and model the process by which they were formed.

\subsection{Initial Observations: Name Molds}

The multi-word phrases used to create names are not all different.
In fact, many different names fall into the same pattern, with only slight variations.
These patterns can be viewed as molds into which a chosen concept word is embedded (which contributes to answering \ref{rq:struct}).
Thus the multiplicity of different names is largely the result of using multiple molds in combination with multiple concept words.

\begin{table}\centering
\caption{\label{tab:combo}\sl
The number of instances of names for the constant 4 (maximal benefits per month) in the benefits card scenario, where names are presented as combinations of using a mold with a concept word.
In the molds, X stands for the concept word (benefit, treat, etc.), which could appear in singular or plural.}
\begin{tabular}{l|c@{}c@{}c@{}c@{}c@{}c@{}c@{}c@{}c@{}c}
\hline
\emph{Mold} & \rotatebox{65}{benefit} & \rotatebox{65}{treat} & \rotatebox{65}{gift} & \rotatebox{65}{bonus} & \rotatebox{65}{perk} & \rotatebox{65}{point} & \rotatebox{65}{reward} & \rotatebox{65}{prize} & \rotatebox{65}{pamper} & \rotatebox{65}{pinuk} \\
\hline
X               & 2     & 1   & 1   & -   & -   & -   & -   & -   & -   & -   \\
max\_X          & 16    & 2   & 1   & 1   & 2   & -   & -   & 1   & -   & 6   \\
max\_X\_per\_month& 13  & 2   & 3   & 1   & -   & 1   & 1   & -   & 1   & 1   \\
X\_per\_month   & 2     & -   & -   & -   & -   & -   & -   & -   & -   & -   \\
max\_monthly\_X & 3     & 2   & 1   & -   & -   & -   & -   & -   & -   & -   \\
max\_month\_X   & 1     & -   & -   & -   & -   & -   & -   & -   & -   & 1   \\
max\_X\_num     & 3     & -   & -   & -   & -   & -   & -   & -   & -   & -   \\
X\_max\_num     & 1     & -   & -   & -   & -   & -   & -   & -   & -   & -   \\
max\_number\_of\_X& 1   & 1   & -   & -   & -   & -   & -   & -   & -   & -   \\
max\_num\_of\_X & 2     & -   & -   & -   & -   & -   & -   & -   & -   & -   \\
max\_X\_amount  & 1     & -   & -   & -   & -   & -   & -   & -   & -   & -   \\
max\_acc\_X     & 1     & -   & -   & -   & -   & -   & -   & -   & -   & -   \\
max\_allowed\_X & 1     & -   & -   & -   & -   & -   & -   & -   & -   & -   \\
monthly\_X\_limit & 1   & 1   & -   & -   & -   & -   & -   & -   & -   & -   \\
\hline
\end{tabular}
\end{table}

An example is given in Table \ref{tab:combo}.
This shows the molds that were identified in the names suggested for the constant 4 in the benefits card scenario (both Hebrew and English versions).
This constant represents the maximal number of benefits that may be accrued in any single month.
For each mold, the number of times it was used with each concept word is shown.
The most popular molds by far were \code{max\_X} and \code{max\_X\_per\_month}, and the most popular concept word was ``benefit''.
And while the combinations of these molds and concept word dominate the table, another 26 combinations were used by at least one subject.

Note, however, that the molds themselves can also be viewed as containing additional concepts.
This leads to a multi-dimensional view of the names instead of the 2D view of Table \ref{tab:combo}.
In particular, each of these names can be viewed as being based on a combination of choices from the following dimensions:
\begin{itemize}
\item \emph{Object}: we are concerned with credit card benefit points.
	This is what we identified as the concept word above.
    Many different words were used in our experiment, with a strong accessibility effect related to the description.
\item \emph{Operation}: we are counting the number of points accrued.
	This is reflected by words like \code{number} (or \code{num}) or \code{acc} (accumulated).
\item \emph{Qualifier}: specifically, this constant concerns the maximal number of points.
	As this is the central function of this constant, nearly all names included the word \code{max}.
\item \emph{Time}: the counting is reset periodically.
	This is reflected by including \code{month} or \code{monthly}.
\end{itemize}
A closer observation shows that these dimensions do not all have the same standing.
The ``object'' and ``qualifier'' dimensions are nearly \emph{universal}, being present in nearly all the names.
But the ``operation'' and ``time'' dimensions are \emph{alternates}: each of the observed name molds used either one or the other.
Thus is appears that developers differed in their opinions on whether it was worth while to extend the name and include these concepts.

The above analysis is not unique to this specific example, and in fact we claim it is fairly common (see Section \ref{sect:model} below).
Another example is the following.
The scenario presented was of a mouse looking for cheese in a maze, where the location of the cheese was changed each day.
The question was to name the variable representing the location of the cheese today.
The answers can be dissected according to the following dimensions:
\begin{itemize}
\item \emph{Location}: this is obviously the main concern of this variable.
	It was also the dimension with the widest variability, employing the words \code{location} (or \code{loc}), \code{coordinates} (or \code{coord}), \code{position} (or \code{pos}), \code{place}, \code{target}, \code{room}, and \code{index}.
\item \emph{Object}: a very focused dimension, with only one concept word, \code{cheese}.
\item \emph{Time}: a dimension with 3 main representations: \code{current} (or \code{cur}), \code{today}, and \code{daily}.
\end{itemize}
In this case, all 3 dimensions were nearly universal.

\subsection{Analyzing Name Structures}

Given the observation that names are composed of words representing different concepts arranged in molds, we wanted to analyze the number of concepts involved in each name and the diversity of words used to represent each concept.
To collect this data we designed an interactive tool that helps in the manual analysis of name structures.

The tool allows the analyst to select the question being analyzed, and loads all the names given as answers to this question.
The names (normalized to lower-case with words separated by \_) are displayed in the rows of a table.
The analyst can then add columns to the table to represent different concepts that appear in the names.
For each name, the analyst identifies words representing different concepts, and copies them to the appropriate columns (creating new columns if needed).
The tool then scans all the following names, and if these words appear in them too it automatically classifies them in the same way.
Thus the manual work is much reduced, and each word has to be classified into a concept only once.
The tool then tabulates the distribution of the number of concepts used in the variable names, and the distributions of the words used for each concept.

The classification of words to concepts was performed by two analysts independently.
In cases of disagreement a third analyst selected which classification to use.
Finally, a fourth analyst (the PI) reviewed all the classifications.
Such a manual oversight was thought to be needed because a mechanical classification into words may miss their meaning in context.
For example, when naming the pay rate for overtime work, two names that were suggested were \code{overtime\_hourly\_rate\_addition} and \code{additional\_hours\_rate}.
We classified the \code{additional\_hours} phrase from the second as belonging to the same concept as \code{overtime} from the first, and further noted that the second name did not contain words representing the concepts of ``additional pay'' and ``per hour'' (as a qualifier of ``rate'').
This is contrary to a mechanistic identification of ``hourly'' with ``hours'' and of ``addition'' with ``additional''.

Note that essentially the same word may appear in different forms, e.g.\ abbreviated or in plural.
When classifying words one has to decide whether to retain this variability or not.
We decided to focus on the semantic level and not on the lexical level, so we normalized all variants of the same word to one basic form.
In addition, if answers included more than one optional name, we used only the first one.

\begin{figure}\centering
\includegraphics[width=0.9\columnwidth]{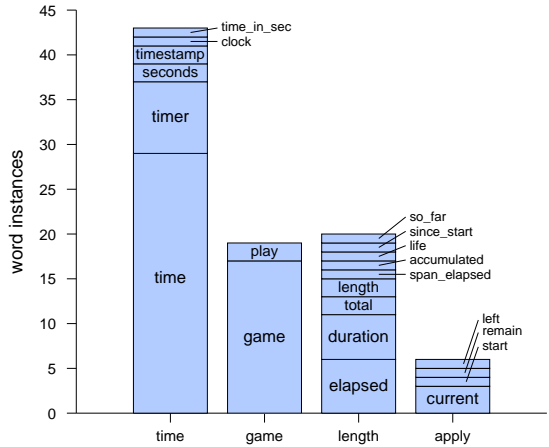}
\vspace*{-4mm}
\caption{\label{fig:game-time-words}\sl
Words used for different concepts in a variable representing the time playing minesweeper (Hebrew description, out of 51 responses).}
\end{figure}

An example of the results is shown in Fig.\ \ref{fig:game-time-words}.
The context is the well-known minesweeper game, and the variable being named represents the playing time.
The most popular concept included in the names was ``time'', and the most popular word to represent it was also ``time'', but several other words were also used.
Other concepts were the fact that we are timing a ``game'', and an emphasis on the fact that we are interested in the ``length'' of time.
A few also noted what the time applied to, usually that it was ``current''.

Tables with identified concepts and words representing them are included in the experimental materials.

\subsection{Fashioning a Model}

To model the process of selecting (or rather inventing) names we pored over the above results, and noted the following phenomena:
\begin{itemize}
\item The repeated use of certain molds
\item That different molds may contain the same concepts
\item The repeated use of specific words to represent concepts in the same or across different molds
\end{itemize}
We then asked what process could lead to these observations?

Taking the results into account, we believe the following conceptual operational model is a good starting point for understanding how developers choose names.
We suggest that name selection involves three choices:
\begin{enumerate}
\item Decide what concepts should be embedded into the name.
\item Decide what word to use to represent each concept.
\item Decide on the structure of the name and the order of the chosen words.
\end{enumerate}

The decision of what \textbf{concepts} to include can be formulated as the identification of the dimensions that should be characterized.
These dimensions are case specific, and the decision on which dimensions to include is perhaps the major decision in naming.
The main consideration is to include information regarding the intent behind using this variable --- what information it holds, and what it is used for.
As a practical matter, if it is felt that a comment is needed to explain the variable's objective, wording from the comment should probably be included in the variable name.
In certain situations it may be prudent to also include an indication of what \emph{kind} of information this is, e.g.\ that a length is in the horizontal or rather in the vertical dimension, or that a buffer contains user input and should therefore be considered unsafe%
\footnote{This is a variation on Hungarian notation; see https://www.joelonsoftware. com/2005/05/11/making-wrong-code-look-wrong/.}.
After the input is checked, it can be stored in another variable, with a name indicating that it is safe.

The second decision is what \textbf{word} to use to represent each dimension.
In many cases some dimensions are very focused, with one specific word being the obvious choice.
This can be the result of simplicity or of an accessibility effect, where the vocabulary used to describe the situation guides developers to use the same terms.
But in our experiments there was often at least one dimension that was highly variable, with many different contending words.
Such diversity may cause problems down the road, if developers become confused whether synonyms actually mean the same thing or represent nuanced differences.
Agreeing on a project lexicon and avoiding the use of synonyms can help \cite{deissenboeck05};
nuances are better replaced with an additional word providing explicit distinctions.
Additional considerations when choosing words include that they be easily distinguished from each other, not too long, memorable, and pronounceable \cite{jones:cbook}.
Adhering to such suggestions makes the resulting names easier to work with.

The decision regarding the \textbf{structure} of a name can be formulated as selecting a mold.
A possible consideration is to follow (English) language rules, e.g.\ that adjectives come before nouns (which would suggest \code{max\_points} over \code{points\_max}).
One can perhaps even add a preposition to make the name into a phrase \cite{liblit06}.
Another obvious consideration is to abide by a project's naming conventions.

The model does not necessarily imply that these 3 steps are cognizant choices made by the developers, nor that they are done serially one after the other.
Thus there may be significant interplay between say the choice of concepts and the choice of a mold.
But the model provides a way to analyze common behaviors by multiple developers in hindsight.
And it explains the commonalities between names given in our experiment.

\begin{figure}\centering
\includegraphics[width=0.7\columnwidth]{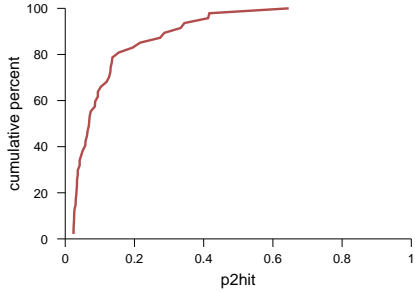}
\vspace*{-4mm}
\caption{\label{fig:p2hit-all}\sl
Cumulative distribution of $p2hit$ for all questions.}
\end{figure}

Conversely, the model also explains the large variability that is observed in names chosen for the same task:
this is a result of multiplying the number of options to make each of the three choices.
When there are many reasonable options, the probability that two developers would select the same name is low.
This is the common case, and indeed the distribution of this probability for all 47 questions listed in Table \ref{tab:2hit} shows that in 80\% of the questions the probability was less than 15\% (Fig.\ \ref{fig:p2hit-all}).
But when there is one dominant concept, and one dominant word to represent it, this probability can become significant.
In our experiments there were two such cases:
using \code{maze} for a data structure describing a maze, and using \code{size} for the field describing the size of a file.

\section{Using the Model}

As noted, we do not claim that the above model is indeed used in a cognizant manner by developers.
But what if it was?
Can the model guide developers in choosing names?
Will the names generated with the help of the model be superior over names that are chosen without it?
A second experiment was performed to answer these questions.
This experiment is essentially a replication of the first experiment, with the difference that subjects are first introduced to the model and asked to use it when they choose names.

\subsection{Research Questions}

More formally, the research questions for the second experiment were
{\renewcommand{\theenumi}{RQ\arabic{enumi}}
\begin{enumerate}
\setcounter{enumi}{\value{rqnum}}
\item\label{rq:model}
Does being cognizant of the name formation model affect the names produced by developers?

\item\label{rq:quality} 
Are names produced with the model superior over names produced without it?
\end{enumerate}}

\subsection{Methodology}

\subsubsection{Changes to the Survey}

The goal of the second experiment was to check the possible effect of using the naming model on the actual choice of names.
To accomplish this we first introduce the subjects to the model.
This is done by providing a brief description of the 3 steps, followed by an example of using it.
The example involved a new scenario --- creating a display to show scores in a bowling alley --- that was not part of the original survey.
The example concerned naming a variable holding the display shown to experienced players, and included a comment about the option of dropping one of the concepts leading to a different name.

In the interest of being able to compare the results, the questions used in the second experiment were identical to those in the first experiment.
But given that the focus is on giving names and not on understanding names, we used only 7 of the original 11 scenarios.
5 of these scenarios had English and Hebrew versions, and the other 2 had 2 English versions and a Hebrew version.
While we were interested only in the questions regarding naming (and not those about understanding), we decided to retain all the questions as in the original experiment.
The reason was that answering seemingly redundant questions can nevertheless affect how the naming questions are answered.
Each scenario was preceded by a 2-line reminder of the steps in the model.

A small pilot study with 3 subjects indicated that using the model is time consuming, with subjects actually using paper to write down concepts and words.
We therefore toned down the wording urging subjects to use the model, and reduced the number of scenarios given to each subject to 3.
As in the first experiment, the scenarios were chosen at random, and subjects were never given more than one version of the same scenario.

The survey was conducted again using the Qualtrics platform, in May 2019 (one year after the first experiment).
100 subjects participated.
62\% of  the  respondents  were  male,  and  34\%  female.
The mean age was 25.4  years, the average programming experience was 3.6 years, and 77\% were students.
The full listing of the survey, including the introduction of the model, is available in the experimental materials.

\subsubsection{Assessing the Quality of Names}

After collecting the results of the survey, the structure of the chosen names was analyzed as in the first experiment.
But to answer \ref{rq:quality} we need to assess the quality of the names given in the second experiment relative to the names given in the context of the same scenarios in the first experiment.
To do this we devised the following protocol.

\begin{enumerate}
\item First, we recruited two 3rd year students to serve as external judges.
We required two in order to compare their judgments with each other.
Importantly, the judges did not know about the experiment or the model, thus eliminating the danger of unintentional bias.
The judges worked independently, but performed exactly the same tasks.
\item The judges judged all 7 scenarios.
For each scenario, they first read the description of the scenario to learn the required background.
\item For each question in the scenario,
    \begin{enumerate}
    \item The judges read the question to understand what was required.
    \item Each judge independently reviewed 60 pairs of variable names, and chose which one they thought was the better name, based on their familiarity with the question and their experience in programming.
    In each pair one name was from the first experiment and one from the second experiment, selected at random uniformly with repetitions from all the names given by subjects.
    Pairs that happened to be the same name were discarded.
    The order of the names in each pair was randomized.
    \end{enumerate}
\end{enumerate}
We used 23 questions in total from all the scenarios together.
Viewing and judging 60 pairs of variable names for each question led to a sum total of 1380 pairs, of which 24 were removed as in retrospect we found they included non-names that should have been excluded.
Reading the scenarios and questions and judging all these pairs took about 7--8 hours of work.
The judges were paid 500 NIS for their effort (approximately \$140).

The end result was a dataset of 60 pairs of names for each of the 23 questions, with the following attributes:
\begin{enumerate}
    \item Which name came from the original experiment and which from the experiment with the naming model.
    \item Which name was presented first to the judges.
    \item The judgments of the two judges as to which name was better.
\end{enumerate}

\subsection{Results}

\subsubsection{Name Structure}

To answer \ref{rq:model} we need to compare the names produced by subjects in the second experiment with the names produced by subjects in the first experiment.
This is done on a question-by-question basis.
Our main concern is with the structure of the names, and specifically, the number of concepts that subjects chose to include.

Naturally, not all subjects used the same number of concepts (or the same specific concepts) in their names.
In particular, a few subjects chose to include unique concepts that nobody else thought were important.
And the more subjects you have, the higher the probability that someone will add a new concept.
As the number of subjects was different in the two experiments, we needed to normalize the results.

\begin{figure}\centering
\includegraphics[width=0.7\columnwidth]{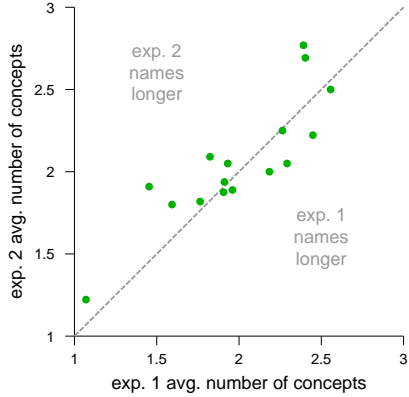}
\caption{\label{fig:cmp-concepts}\sl
Comparison of the average number of concepts used in variable names in questions with Hebrew descriptions in the two experiments.}
\end{figure}

We perform the normalization by repeated random sampling from the larger group.
Let $N_1$ denote the number of respondents in the first experiment, and $N_2$ in the second.
Assume $N_1 > N_2$.
for experiment 2, we simply report the average number of concepts used by the $N_2$ respondents.
But for experiment 1, we use a bootstrap-like procedure.
We sample $N_2$ respondents out of the $N_1$ available respondents, and calculate the average number of concepts used by this sample.
We then repeat this 100 times.
The average of the 100 repetitions is taken as the average number of concepts in experiment 1.

The results are shown in Fig.\ \ref{fig:cmp-concepts}.
In this scatter plot, each dot represents a question.
The dots coordinates are the average number of concepts used in the requested variable name (corrected for sample size as described above).
The results indicate that in the second experiment, when subjects were coached on using the model, they tended to use names with more concepts.
Focusing on the questions with Hebrew descriptions, this happened in 10 cases and the names were 0.20 concepts longer on average.
In 6 other cases the names from the first experiment were longer, by an average of 0.14 concepts.

\subsubsection{Name Quality}

\begin{figure}\centering
\includegraphics[width=0.9\columnwidth]{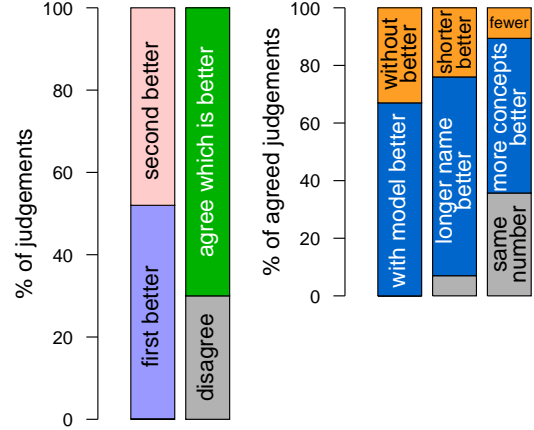}
\caption{\label{fig:judges}\sl
Classifications of the results of judging the quality of variable names.}
\end{figure}

We now turn to \ref{rq:quality}.
The results of judging the quality of variable names are shown in Fig.\ \ref{fig:judges}.
The left-most bar shows that the judges were largely unaffected by the order in which names were presented: they preferred the first name 52\% of the time, and the second name in the remaining 48\%.
Given the large number of judgments performed this deviation from a 50--50 split is large, but it is still not statistically significant (p=0.075 using binomial test).

The next bar shows that in 70\% of the trials the two judges agreed on the preferred name.
In 30\% they disagreed.
Using Cohen's kappa to assess the degree of agreement yields $\kappa = 0.366$ which is considered a fair level of agreement.

Focusing on the 950 cases in which the judges agreed, in 67\% of them they judged the name chosen by subjects using the model to be superior.
Thus names chosen when explicitly using the model were favored over names that were chosen with no such framework by a ratio of 2:1.
This and the following results are all highly significant (p<0.0001).

When looking at the characteristics of the names, we found that longer names were preferred in 74\% of the cases in which there was a difference in length (in 7\% the names were the same length).
When considering concepts, in 36\% of the cases both names had the same number of concepts.
But of the cases where there were different numbers, the names with more concepts were preferred in 83\% of the cases.
These results agree with the previous results on the effect of using the model.
They indicate that the model helps by encouraging subjects to create longer names, and to do so by including more concepts.
Note that this is different from creating longer names just by being verbose.

\subsubsection{Subjective Opinion}

At the end of the survey we added two questions regarding the model.
The first simply asked whether the model was used by the subject taking the survey.
The second asked whether the subject felt that the model helped in choosing names.
The results were that 52 respondents reported that they used the model and 16 reported that they did not.
Of those that used the model, 35 reported that it helped and the remaining 17 that it did not.

Using the data from respondents who claimed not to have used the model implies that the results presented above regarding the benefits of the model should be considered as conservative.
We did not exclude such respondents from the analysis to avoid reducing the number of samples, and the issue of what to do with respondents who did not answer this question.
We further note that even if they did not use the model explicitly, reading about it might have affected their work.

\section{Threats to Validity}

Construct validity refers to the degree to which we can be sure we are measuring what we set out to measure.
One of our goals was to measure spontaneous name choice, but this may suffer from bias due to accessibility.
We therefore used Hebrew descriptions to avoid enhancing the accessibility of English words.
However, bilingual subjects may still translate the Hebrew terms so some accessibility effect may remain.
This may also be affected by subjects' differences in English language knowledge.
Lack of sensitivity to English nuances can also affect the understanding of names.
For example, we speculate that native English speakers would not think that in a call to \code{resize(factor)} the parameter is the new size.

Another threat exists in analyzing the structure the names.
We used a conservative approach to splitting a name into words, which may miss some cases \cite{hill14}.
A harder problem is identifying concepts.
For example, do the terms \code{full\_time} and \code{normal} refer to the same concept in the context of working hours?
We used a protocol in which 2 analysts independently coded the structure, another reconciled differences of opinion, and a fourth reviewed the results.
Still, other analysts may prefer a different coding and this may lead to different results.
This can be resolved only by replication studies.

In addition, using an online survey may also lead to threats to construct validity.
Respondents to a survey may answer immediately without thinking it through, whereas when actually programming they may spend more time to consider things.
Also, they may skip the longish story introducing the questions, and guess the context from the questions themselves.

Internal validity is concerned with causation, i.e.\ whether the results can indeed be attributed to differences between treatments.
In the first experiment, we are mainly concerned with variation emanating from differences between subjects, so the issue is moot.
However, differences between treatments also exist, e.g.\ when descriptions are given in Hebrew as opposed to in English.
One threat in this context is that differences in English proficiency may also affect the results.

In the second experiment, we change the treatment by introducing the model of name formation.
Except for this difference, the experiment is identical to the previous one, so differences are indeed expected to be an outcome of this change.
However, we can not be sure that the differences are due to using the model we had suggested.
It is possible that by suggesting the model we caused subjects to think more about the naming, but the model itself had no effect.
Settling this threat would require a new set of experiments, as mentioned below under suggestions for future work.

External validity refers to the generalizability of the results, namely whether they can be expected to hold beyond the specific experimental conditions that were used.
A possible issue is the use of bilingual subjects, in our case native Hebrew speakers for whom English is a second language.
While this enabled us to mitigate the accessibility effect, it exposes us to the risk that the results do no represent the behavior of native English speakers.
On the other hand, we note that many developers in the world are not native English speakers.

Another well known risk to external validity is the question of the experience of the subjects, and, specifically, whether students can be used in lieu of professional developers.
This concern was mitigated to some degree by using both, and ensuring that these labels indeed reflected differences in experience.
While we presented initial results showing that naming is affected by experience, this issue certainly deserves additional work.

In a related vein, we only used a modest number of example scenarios, and found a diverse set of results.
We make no claim that our scenarios are necessarily representative, and cannot infer any statistics on how common they are.
Such issues require extensive replications and variations by other researchers in different contexts \cite{feitelson15}.

Finally, a major question concerns the validity of our model.
This is merely a conceptual model, and our only claim is that it may be useful in understanding and perhaps also in guiding naming.
At the same time, other models are certainly possible.

\section{Conclusions}

\subsection{Summary of Results}

Returning to our research questions, our results suggest the following answers.
For \ref{rq:same} the probability that different developers would suggest the same name in the same situation is very low.
Our analysis suggests that this divergence is the result of a multiplicative combination of possible choices at the three steps of name creation: deciding what concepts to include in the name, how to represent each concept, and how to structure it all together.

Leading up to this result, for \ref{rq:prime} we found that the choice of words can be strongly affected by the wording used in describing the scenario.
A practical implication of this finding is the importance of using a consistent vocabulary, and perhaps even the desirability of maintaining a project lexicon to guide programmers towards the use of mutually-agreed names \cite{caprile00}.
In \ref{rq:exp}, we also found that experienced programmers tend to choose slightly longer names.

For \ref{rq:understand}, we found that the divergence in choosing names does not necessarily harm one's ability to understand existing names.
But there are cases where names are misunderstood, or when assumptions regarding technical details differ.
It is hard to anticipate when this will happen.
Thus, the only way to find and improve problematic names is to check.
This implies that developers should include naming issues explicitly in code reviews:
make sure reviewers understand names the same way, and make the same assumptions about the named entities.

Regarding \ref{rq:struct}, we found that names tend to be multi-word with an average of about 5 letters per word.
Analyzing the words used showed that they can typically be classified into separate concepts.
This led to the model of name formation, in which the first two steps are to decide on the concepts to include in the name and on the words that will represent each concept.

\ref{rq:model} took the 3-step model of naming as its starting point, and asked whether coaching programmers on using this model would affect the variable names they choose.
The results were that indeed such an effect exists, and respondents who were coached with the model tended to use longer names with more concepts.
In \ref{rq:quality}, we then found that these names were generally also judged to be superior over names chosen by respondents who had not been exposed to the model.

\subsection{Future Work}

While the reported experiments provide interesting new information about how names are chosen and how the process of choosing names may be improved, they also suggest several venues for future work.

The most immediate are experiments to verify whether the suggested model was indeed instrumental in creating better names.
For example, one could conduct an experiment in which subjects are just asked to spend time thinking about names, without any mention of the model.
Alternatively, one can instruct the subjects explicitly on the advantages of using longer names with more concepts.
If these experiments also produce superior names, this may suggest easier ways to affect naming by practitioners.

Assuming the model is accepted, additional experiments can be designed to try to influence each step.
However, these would probably require work on a larger scale on real projects, and not just on scenarios.
Possible ideas include:
\begin{itemize}
\item Providing project-level documentation, and checking whether the explicit identification of central concepts encourages use of these concepts in names.
\item Providing a project vocabulary and checking how it affects word choice in names.
\item Checking the effect of guidelines for name structures on name choices and on name consistency.
\end{itemize}

Digging deeper, we note that naming may depend on mindset.
In a question about naming a variable that represents the state of an elevator door, one answer included the comment that, if it was a Boolean, the name would be \code{is\_open}, but, if it was an enumerated type, the name would be \code{door\_state}.
Hence understanding the considerations that go into naming requires more detailed information to be collected.
This can be done with think-aloud studies, where subjects are asked explicitly to explain why they chose each name, or by using follow-up questions asking about the rationale for names after-the-fact.

A related issue is that today many software developers are not native English speakers, but programming --- and naming --- are done in a predominantly English setting.
Non English speakers may choose sub-optimal names because they are unaware of certain words' connotations.
Assessing the prevalence of such problems and suggesting ways to mitigate them are important research directions.

An especially intriguing issue is the relative importance and quality of names: which of them convey the most meaning.
This has been investigated in the context of code summarization tasks, where developers have been observed to use names in their summaries \cite{haiduc10,rodeghero15}.
It would be interesting to see whether the selected names have any structural features in common, and whether they can serve to educate how to instill meaning into names.

Our study is thus just a first step.
We achieved a superficial description of name structure, and a possible model of how names are formed.
We also showed that it is possible to affect and improve the process of naming.
These results should be followed by research on the interaction of naming with cognitive processes in the brain \cite{storey06}.
For example, how are names understood, and why are names with more concepts better?
And all told, do longer names increase the cognitive load (you need to remember more) or reduce it (by providing more information)?
Answering such questions will provide a scientific basis on which software engineering practices can be built.

\section*{Experimental Materials}

The experimental materials and results are available at
\code{http://bit.ly/names2019}.

\bibliographystyle{myabbrv}
\bibliography{abbrv,misc,se}

\section*{Appendix: The Survey Questions}

\renewcommand{\thesubsection}{\arabic{subsection}}

\subsection{Minesweeper}

Minesweeper is a known simple game defined as follows:
\begin{itemize}
\item The player is initially presented with a grid of undifferentiated squares.
\item Some randomly selected squares, unknown to the player, contain ``mine''.
\item The game is played by revealing squares of the grid by clicking them. One of the following will happen:
\begin{itemize}
    \item If a square containing a mine is revealed, the player loses the game.
    \item If no mine is revealed, a digit is displayed in the square, indicating how many adjacent squares contain mines.
    \item If there are no adjacent mines, a set of squares is revealed - all the empty squares until (and including) the boundary with numbered squares.
\end{itemize}
\item The game purpose is to reveal all mine-free squares in the shortest time.
\end{itemize}
\begin{enumerate}
\item Are you familiar with this game?
\item The game's level of difficulty depends on the grid size and the number of mines in it.
Write a function signature for a function which receives the above parameters and returns the level of difficulty of the game.
\item How would you call the variable which holds the game time?
\item We will have a data structure which assigns a number to each square in the board as follows:
\begin{itemize}
\item -1 if the tile contains a mine
\item The number of mines in the adjacent squares otherwise
\end{itemize}
How will you name this data structure?
\item What do you think is the function of the following interface: \code{expose(row, col)}?
\end{enumerate}

\subsection{Salary}

In a large chewing gum company, workers earn hourly (NIS).
Every employee has a fixed hourly wage value.
\begin{enumerate}
\item Given the following interface: \\
\code{pay (hours, rate)} \\
What do you think the interface does?
\item What is the purpose of hours parameter?
\item What is the purpose of rate parameter?
\item Implement the interface.
\end{enumerate}
Purim is right around the corner and the Mishlochei Manot cause an increased demand for chewing gum.
To overcome this the factory manager encourages employees to work overtime as follows:
\begin{itemize}
\item A full-time position requires 45 weekly work hours.
\item After 45 weekly work hours, the hourly wage for the employee increases by 10 ILS.
\end{itemize}
To implement this some variables were added:
\begin{itemize}
\item A constant containing the value 45
\item A variable for the hourly wage during overtime pay
\end{itemize}
\begin{enumerate}
\item Name the constant containing the value 45.
\item Name the variable for the hourly wage during overtime pay.
\end{enumerate}

\subsection{Maze [alternative version with ``labyrinth'']}

Columbus the mouse lives in a maze, in which every day cheese is placed at a different location and he would like to find out where is the cheese hidden. Columbus is a pedantic mouse, and so he traverses all the rooms in an orderly fashion. Specifically he does not re-enter a room which he already visited.

Assuming you will be asked to write a program for Columbus' algorithm,
\begin{enumerate}
\item How would you call the variable holding the location of the cheese today?
\item How would you call the variable (or data structure) keeping track of where has he already visited?
\item How would you call the data structure describing the maze?
\end{enumerate}

\subsection{Tic-Tac-Toe [alternative version with ``game grid'']}

Assuming you would need to write a program for playing Tic-Tac-Toe,
\begin{enumerate}
\item How would you name the variable (or data structure) describing the current state of the game board?
\item The implementation includes a function \code{makeTurn(int row, int col)}. What do you think it does?
\item What is the purpose of the ``row'' parameter?
\item What is the purpose of the ``col'' parameter?
\item Playing requires displaying the board to the user.
Propose a function signature for this purpose (function name + parameters).
\end{enumerate}

\subsection{File Management}

A computer program manages files.
Files can be added or deleted, and for each file, its size and name are stored.
Assuming you would need to write a program implementing the file manager,
\begin{enumerate}
\item The implementation contains a class which describes a file within the system.
How would you name the field describing the file's size?
\item The implementation contains a function \code{arrangeFilesByName(files)}.
In your opinion, what does it do?
\item What is the role of the parameter? what is its type?
\item What do you think it returns?
\item The implementation includes a function that receives a file, by how much we want to increase it, and the available space on the disk, and checks whether there is enough space.
Suggest a signature for this function.
\end{enumerate}

\subsection{Ice Cream Sandwich}

Summer is coming and Ori is planning to make some money during his break.
Ori loves ice cream and has a great ice cream sandwich recipe. To make one sandwich the following 3 ingredients are needed:
\begin{itemize}
\item 2 chocolate biscuits
\item Half a cup vanilla ice cream
\item 20x10 cm wrapping paper
\end{itemize}
\begin{enumerate}
\item Write an API function signature to help Ori calculate, given the quantities of ingredients he has, how many sandwiches he can produce.
\item What do you think
\code{profit (units, cost, price)}
function does?
\item What is the purpose of the ``units'' parameter?
\item What is the purpose of the ``cost'' parameter?
\item What is the purpose of the ``price'' parameter?
\end{enumerate}

\subsection{Elevator}

The elevator in the CS building has broken. Parts of the control system's code were deleted and now students were asked to re-implement them.
You have the following functions available for use:
\begin{itemize}
\item Open door.
\item Close door.
\item Go down a number of floors.
\item Go up a number of floors.
\end{itemize}
as well as a field indicating the current position of the elevator.

Following is a piece of code implemented by Stav:
\code{\begin{tabbing}
xxxx\=xxx\= \kill
if (var1>var2) \\
\> direction = "Up" \\
\> var3 = var1 - var2 \\
\> goUp(var3) \\[2mm]
if (var1<var2) \\
\> direction = "Down" \\
\> var3 = var2 - var1 \\
\> goDown(var3)
\end{tabbing}}
The original variable names have been replaced by var1, var2, var3.
\begin{enumerate}
\item What does this code do?
\item Replace \code{var1} with a name you would have used.
\item Replace \code{var2} with a name you would have used.
\item Replace \code{var3} with a name you would have used.
\item How would you call the variable describing the state of the elevator's door (open/closed)?
\item Stav wants to check whether the code was properly corrected.
She goes into the elevator on the 0 ground floor and presses all the buttons together.
It is expected that the elevator will rise and open at each floor.

Write a pseudo code loop that passes all the floors in ascending order from floor 0 and opens the door on each floor.
\end{enumerate}

\subsection{Benefits Card}
``Benefits card'' is a credit card company which allows its customers to accumulate benefits which can be exchanged for various offers.
Benefits are accumulated as follows:
\begin{itemize}
\item Benefits are accumulated once per month.
\item A customer is entitled to 1 benefit for each 2,000 ILS which are billed to the credit card during the previous month.
\item Up to 4 benefits can be accumulated per month.
\item Benefit entitlement does not cross over to the next month.
Benefits not used during the month given will expire.
\end{itemize}

\begin{enumerate}
\item Name the constant holding the value 4 according to its purpose.
\item Name the constant holding the value 2000 according to its purpose.
\item Name the variable holding the number of benefits the client is entitled to during the current month.
\item When a customer wants to use a benefit, the system executes a function which works as follows: \\
Input:
\begin{itemize}
\item Number of benefits the customer is entitled to during the current month
\item Number of benefits the customer has used during the current month
\end{itemize}
The function calculates the difference, and returns TRUE if the balance is positive.

Write a function signature for this function.
\end{enumerate}

\subsection{Rock-Paper-Scissors [Hebrew only]}

You want to implement the game Rock-Paper-Scissors.
The implementation includes a function \code{whoWins(playerA, playerB)}.
\begin{enumerate}
\item What do you think is the type of the return value?
\item What are possible values of the return value?
\item What are the roles of the parameters?
\item What are their types and possible values?
\end{enumerate}

\subsection{Add [Hebrew only]}

A library contains the following function: \code{Add(a, \ b)}. 
What do you think is the return value of \code{add([1,2,3], [4,5,6])}?

\subsection{Resize [Hebrew only]}

In an image processing library there is a function \code{resize(factor)}. 
\begin{enumerate}
\item What do you think the function does?
\item An image object has a field specifying the image width.
Write the line of code that updates this field in the above function.
\end{enumerate}

\end{document}